\documentclass[prd,twocolumn,aps,superscriptaddress]{revtex4}
\usepackage{amsfonts}

\usepackage{dcolumn}
\usepackage{bm}
\usepackage{tikz}
\usepackage{graphicx}
\usepackage[colorlinks=true,citecolor=blue,urlcolor=blue]{hyperref}
\usepackage{amsmath,amsfonts,amssymb,times,natbib}
\usepackage[standard]{ntheorem}

\begin{document}

\title{ Amplification of Gravitationally Induced Entanglement }

\author{Tianfeng Feng} 

\affiliation{State Key Laboratory of Optoelectronic Materials and Technologies and School of Physics, Sun Yat-sen University, Guangzhou, 510275, People's Republic of China}

\author{Vlatko Vedral} 
\affiliation{Clarendon Laboratory, University of Oxford, Parks Road,
	Oxford OX1 3PU, United Kingdom}
\affiliation{Centre for Quantum Technologies,	National University of Singapore}
\affiliation{Department of Physics, National University of Singapore}

\date{\today}
\begin{abstract}
 Observation of gravitationally induced entanglement between two massive particles can be viewed as implying the existence of the nonclassical  nature of gravity.
However, weak interaction in the gravitational field is extremely small so that gravitationally induced entanglement is exceptionally challenging to test in practice.
 For addressing this key challenge, here we propose a criterion based on the logical contradictions of weak entanglement, which may boost the sensitivity of the signal due to the gravitationally induced entanglement.  Specifically, we make use of the weak-value scenario and Einstein-Podolsky-Rosen (EPR) steering.  We prove that it is impossible for a classical mediator to act on two local quantum objects  to simulate amplified-weak-value phenomenon in two-setting EPR steering.  Our approach can amplify the signal of gravitationally induced entanglement that were previously impossible to observe, by any desired factor that depends on the magnitude of the weak value.  Our results not only open up the possibility of exploring nonclassical nature of gravity in the near future, but also pave the way for weak entanglement criterion of a more general nature.


\end{abstract}

\maketitle



\section{Introduction}
Quantum theory and general relativity, the two backbones of modern physics, have been verified with very high precision in their respective fields. Yet, it is hard to unify them into a unique corpus of laws. One possible route to that general theory is the quantisation of gravity, with the same spirit as other field theories. However, there is a long-standing debate whether gravity should be quantized \cite{Witt, Witt2, Peres, Eppley,CM}.  
 Traditionally, it is believed that the effects of quantum gravity should occur at high energy scales or in the short length regime which are beyond the reach of current technology. 
  Recently, there has been a revival of the idea of a tabletop probe, 
  which highlights the interaction of the probe mass with the gravitational field generated by another mass~\cite{Feynman,Gorelik,Hu1,Hu2,CMVV}.
Especially, two gravity-induced-entanglement tests, sometimes called the Bose-Marletto-Vedral (BMV) experiments \cite{Bose,VV},  have been proposed, which may be use to expose the quantum nature of gravity.
 BMV's protocol aims to provide a firm evidence on whether the gravitational field is mediated by the transfer of quantum information.  
 Bose \emph{et. al.} suggest that if we admit the central principle of quantum information theory: entanglement between two systems cannot be created by local operations and classical communication (LOCC), then gravitationally induced entanglement indicates that gravity must be quantum \cite{Bose}. On the other hand, Marletto and Vedral argue for this view through a more general information-theoretic argument \cite{VV,Marletto}, which is based on constructor theory \cite{David}. Specifically,  one does not need to assume any specific dynamics law of mediators (in this case the gravitational field) to justify the conclusions of creating entanglement in the experimental proposal \cite{VV,Marletto}. In this article, we will focus on the quantum formalism.  
Till now, a variety of advanced theoretical and experimental proposals have been suggested to investigate the gravitationally induced entanglement and nonclassicality \cite{CVD,Brukner,Bose2,Tanjung,Hall,Chevalier,Kamp,Kent,setup2,setup22,Bose3,Bose4,PRX,PRX2,Miao,Schut,Datta,Rijavec,Marios,CM3, Neppoleon, Bose5, Kanno,Animesh,Wald,Had}.
 
Entanglement witnesses are a suitable method for measuring the gravitationally induced entanglement  \cite{HHH,Bourennane}. 
Unfortunately, due to the extremely weak strength of gravity, a ``strong'' and detectable entanglement signal might require a longer interaction time of massive particles in a superposition of two location (matter-wave-like interferometer), which poses a serious challenge to current experimental techniques. As we all know, a general rule of thumb is that the larger and heavier a particle is, the shorter its coherence time. In particular, the experiments must be implemented within the coherent time otherwise the loss of entanglement due to decoherence would prevent us from concluding anything about the quantum nature of gravity. Is it possible to detect weakly entangled signals with limited coupling time for a given mass of particles and a finite resolution or sensitivity of the measurement devices?  Could we amplify the signals of these non-classical correlations?  This is an issue that has not been mainly considered in previous studies \cite{CVD,Brukner,Bose2,Tanjung,Hall,Chevalier,Kamp,Kent,setup2,setup22,Bose3,Bose4,PRX,PRX2,Miao,Schut,Datta,Rijavec,Marios,CM3, Neppoleon, Bose5, Kanno,Animesh,Wald,Had} and is also the main motivation for our present paper.

There is a famous parametric amplification approach in quantum information field, called weak-value amplification \cite{A,Weak,Weak2}. Weak values have their root in quantum weak measurement, which describes a weak coupled measurements, proposed by Aharonov, Albert, and Vaidman \cite{A}. Weak-value amplification exploits the fact that the post-selection of  the weak measurement of a pointer can yield an amplified shift which is exceptionally sensitive to small changes in an interaction parameter. This has been successfully applied to the estimation of a range of small physical parameters \cite{Weak2}, including beam deflection \cite{Hosten,Dixon}, frequency shifts \cite{Star}, phase shifts \cite{Xu} and so on. 


 
In this article, we propose a criterion for determining weak-gravitationally-induced entanglement, which makes use of  a weak-value scenario and Einstein-Podolsky-Rosen (EPR) steering \cite{EPR,EPR1,EPR2}. 
 Specifically, we unify the weak measurements (weak value amplification scenarios) in the framework of EPR steering. Similar to the Bell  test \cite{Bell,CHSH}, we consider two sets of measurement bases that can be randomly selected, one of which is the normal measurement basis (e.g., the computational basis) and the other one corresponding to weak value amplification.
We present a comparison of two predictions of the quantum and classical mediator, the measurement probability distribution and the measurement visibility.
We show that in the case of weak entanglement, the classical mediator (in this case, the gravitational field) cannot simulate the results related to the measurement visibility of weak-value basis, thus ruling out the separable model. Concretely, our approach can amplify the signal of gravitationally induced entanglement by any desired factor that depends on the magnitude of the weak value. Compared to the previous protocols, our approach allows us to observe entangled signals that were previously impossible to observe.
  Besides, our criterion is not limited to the detection of weak entanglement in gravity. It is applicable to more general case of weak entanglement, including potentially macroscopic entanglement. 

        

\section{Quantum formalism of BMV experiments}
\begin{figure}[tb]
 	\center
 	\includegraphics[scale=0.35]{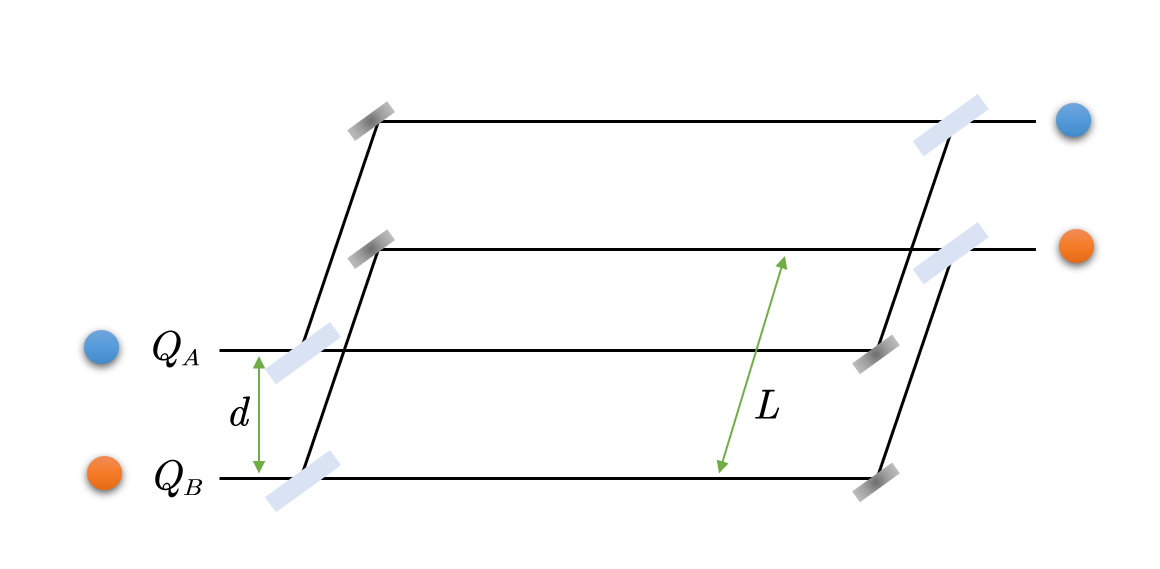}
 	\caption{Symmetric Bose-Marletto-Vedral experiment for testing gravitationally induced entanglement. There are tow mass $\mathcal{Q_A}$ and $\mathcal{Q_B}$.  Each mass  individually undergoes Mach-Zehnder-type interference in parallel, and interacts with the other mass via gravity. 
 	}
 	\label{scheme2}
 \end{figure}

Here we focus on the quantum formalism of BMV experiments.
As  shown in Fig. \ref{scheme2},  BMV proposal is presented. Two quantum mass $\mathcal{Q_A}$ and $\mathcal{Q_B}$ are initially at distance from each other. Each mass individually undergoes Mach-Zehnder-type interference  in parallel, and interacts with the other mass via the gravitational field—which plays the role of the mediator $\mathcal{M}$.  
Under the assumption of locality, observation of gravitationally induced entanglement between $\mathcal{Q_A}\oplus\mathcal{Q_B}$ is the indirect evidence of nonclassicality (quantumness) of the mediator $\mathcal{M}$ \cite{VV,Bose,Marletto}.   Specifically,
 The initial state of system $\mathcal{Q_A}$ and system $\mathcal{Q_B}$ is a separable state (by the first beam splitter),  donated as $\varrho_{A}\otimes\varrho_{B}=|+\rangle_A \langle +| \otimes |+\rangle_B \langle  +|$, where $|+\rangle=\frac{1}{\sqrt{2}}(|0\rangle+|1\rangle$. 
 Since the masses on different paths interact via the gravitational field, the state of the composite system becomes, before they enter their respective final beam splitters,
 $	\varrho_{AB}=\Lambda(\varrho_{A}\otimes\varrho_{B})$,
 where $\Lambda(\cdot)$ is the  map of channel (operation) acting on quantum systems $\mathcal{Q_A}$ and $\mathcal{Q_B}$   induced by the mediator  $\mathcal{M}$. 
 If the quantum state $\varrho_{AB}$ can not be written as  $\sum_i p_i\varrho^i_A\otimes \varrho^i_B$, then $\varrho_{AB}$ is a entangled state, which indicates that the action $\Lambda(\cdot)$ is a entanglement operation. This results may be the evidence of quantumness for mediator $\mathcal{M}$  \cite{Marletto,VV, Wald} .
 On the contrary, a classical mediator can only produce unentangled quantum state for  quantum systems $\mathcal{Q_A}$ and $\mathcal{Q_B}$ , that is 
 $	\varrho^{C}_{AB}=\Lambda^C(\varrho_{A}\otimes\varrho_{B})= \sum_i p_i\varrho^i_A\otimes \varrho^i_B$,
 where  $\Lambda^C(\cdot)$ denotes the effective channel induced by a classical mediator. 
 
 \emph{ Weak-value scenario of BMV experiments.}
Under the time evolution of the joint state of the two masses is purely due to their mutual gravitational interaction  \cite{Marletto,VV}. Thus, the channel $\Lambda(\cdot)$ of quantum mediator $\mathcal {M}$ mentioned above (in Fig.  \ref{scheme2})  is an unitary operation. Specifically, the unitary is given as
$	U=\text{exp}(-i\frac{H\tau}{\hbar})=\text{cos}(\frac{\Delta\phi\tau}{2\hbar})I\otimes I+i\text{sin}(\frac{\Delta\phi\tau}{2\hbar})Z\otimes Z$,
where $H=-\frac{\Delta\phi}{2}Z\otimes Z$ is the Hamiltonian of the two mass is \cite{setup2, setup22}
, in which 
$\Delta \phi=Gm_1m_2(\frac{1}{d}-\frac{1}{\sqrt{d^2+L^2}})$.
After the unitary evolution, the joint quantum state of $\mathcal{Q_A}$ and $\mathcal{Q_B}$ becomes 
\begin{equation} 
		\begin{split}
|\Psi\rangle&=U|+\rangle_A \otimes |+\rangle_B\\
&=\text{cos}(\frac{\Delta\phi\tau}{2\hbar})|+\rangle_A |+\rangle_B +i\text{sin}(\frac{\Delta\phi\tau}{2\hbar})|-\rangle_A |-\rangle_B, 
	\end{split}
\end{equation}
which is a two-qubit entangled state. 
Without loss generality,  $|\Psi\rangle$ can be rewritten in other basis. 
For the qubit case,  the identity can be expressed as $I=|\epsilon\rangle \langle \epsilon|+|\epsilon^{\perp}\rangle \langle \epsilon^{\perp}|$, satisfying $\langle \epsilon^{\perp}|\epsilon\rangle=0$.  Here we define $|\epsilon\rangle=\epsilon|0\rangle-\sqrt{1-\epsilon^2}|1\rangle$ and  $|\epsilon^{\perp}\rangle=\sqrt{1-\epsilon^2}|0\rangle+\epsilon|1\rangle$, where $\epsilon$ is a real positive number. Now the composite state becomes
\begin{equation} \label{gravitystate}
		\begin{split}
	|\Psi\rangle&=(|\epsilon\rangle \langle \epsilon|+|\epsilon^{\perp}\rangle \langle \epsilon^{\perp}|)\otimes I 	|\Psi\rangle\\
	&=\alpha |\epsilon\rangle_A \otimes |\tilde{\chi}_{\epsilon}\rangle_B + \beta |\epsilon^{\perp}\rangle_A \otimes |\tilde{\chi}_{\epsilon^{\perp}}\rangle_B \\
		\end{split}
\end{equation}
where $\alpha= \langle \epsilon|+\rangle$, $\beta=\langle \epsilon^{\perp}|+\rangle$,
$|\tilde{\chi}_{\epsilon}\rangle=	\text{cos}(\frac{\Delta\phi\tau}{2\hbar})  |+\rangle_B+i\text{sin}(\frac{\Delta\phi\tau}{2\hbar})A^{\epsilon}_w  |-\rangle$, and $
|\tilde{\chi}_{\epsilon^{\perp}}\rangle=	\text{cos}(\frac{\Delta\phi\tau}{2\hbar})  |+\rangle+i\text{sin}(\frac{\Delta\phi\tau}{2\hbar})A^{\epsilon^{\perp}}_w  |-\rangle$.
As can be seen, the weak values are embedded in the quantum states$|\tilde{\chi}_{\epsilon}\rangle$ and $|\tilde{\chi}_{\epsilon^{\perp}}\rangle$, which are given as  $A^{\epsilon}_w=\frac{\langle \epsilon|Z|+\rangle}{\langle \epsilon|+\rangle}=\frac{\epsilon+\sqrt{1-\epsilon^2}}{\epsilon-\sqrt{1-\epsilon^2}}$ and $A^{\epsilon^{\perp}}_w=\frac{\langle \epsilon^{\perp}|Z|+\rangle}{\langle \epsilon^{\perp}|+\rangle}=\frac{-\epsilon+\sqrt{1-\epsilon^2}}{\epsilon+\sqrt{1-\epsilon^2}}=-\frac{1}{A^{\epsilon}_w}$ respectively.
From the nature of $|\Psi\rangle$, it follows that if the quantum system $\mathcal{Q_A}$ is projected into $|\epsilon\rangle$ ($|\epsilon^{\perp}\rangle$), then the quantum state of system $\mathcal{Q_B}$ will collapse to (unnormalized) state $|\tilde{\chi}_{\epsilon}\rangle$ ($|\tilde{\chi}_{\epsilon^{\perp}}\rangle$), and vice versa.  The amplified weak-value  $A^{\epsilon}_w$ can be achieved if the result of the collapse of quantum system $\mathcal{Q_A}$  to $|\epsilon\rangle$ when $ \alpha=\langle \epsilon|+\rangle$ is very small. 
From this perspective, the generation of weak value can be explained as it originated from EPR steering \cite{EPR,EPR1,EPR2}, which is determine by the measurements of one of the parties~\cite{note1}.  %


As we have shown above, the weak value $A^{\epsilon}_w$ ($A^{\epsilon^{\perp}}_w$) determine the form of the quantum state $|\tilde{\chi}_{\epsilon}\rangle$ ($|\tilde{\chi}_{\epsilon^{\perp}}\rangle$). The larger the weak value $A^{\epsilon}_w$, the bigger (smaller) the component $|-\rangle$ of the quantum state $|\tilde{\chi}_{\epsilon}\rangle$ ($|\tilde{\chi}_{\epsilon^{\perp}}\rangle$). According to the theory of weak-value amplification, the quantum state $|\tilde{\chi}_{\epsilon}\rangle$  is more likely to be accurately measured with big weak value $A^{\epsilon}_w$ when the phase parameter $\frac{\Delta\phi\tau}{2\hbar}$ is extremely small. So one may use such amplification phenomenon to enhance sensitivity of signal of gravitationalliy induced entanglement.
Unfortunately, the weak-value amplification approach is specific to parametric amplification, and it cannot be used directly to rule out the possibility of classical models. 
Therefore, one need to find an entanglement criterion with weak-value amplification  to exclude the model of classical mediator.

\section{ EPR steering and weak-value scenario} Here we focus on how to construct an entanglement criterion with the weak-value amplification. 
As we have shown above, the weak-value scenario is a special case of EPR steering, which corresponds to one-measurement setting. It is known that the experimental results of one-measurement setting in EPR steering can be easily simulated by a local model. Therefore, from this point of view, the amplified weak value in BMV experiments may be simulated by a classical mediator. That is, one cannot determine the quantumness of gravity directly with a weak value amplification scenario. 
In general, 
EPR steering scenario needs at least two different measurement bases (two-measurement setting) to determine whether the joint quantum state is steerable (entangled) or not. 
Hence, one may consider exploiting the EPR-steering scenario to determine entanglement while keeping the measurement basis corresponding to the weak value amplification as one of the two measurement bases for EPR steering. 

Nowadays, EPR steering has been heavily studied, including the detection of various linear and nonlinear inequalities (see review \cite{Uola}) . There are also some quantum steering paradox based on logical contradictions \cite{Chen,Chen1,Feng}. However, we will show that none of these can be directly used for the verification of  weakly amplified versions of quantum steering. The reason is that all of these depend on the expectation value, which is related to the probability (the probability of weak amplification is very low). This could lead to experimental errors masking the true entangled signal. In the following, we consider not only the probability distribution of the steered quantum states, but also introduce a physical quantity, the \emph{visibility} of the measurement ($\Pi$) of one of subsystem, i.e. $\mathcal{Q_B}$. We express  this quantity in terms of $V=\text{Tr}(\varrho_B \Pi)$, where $\varrho_{B}$ is the steered normalized density matrix of $\mathcal{Q_B}$.  We will show that the genuine entanglement signal is hidden in the visibility. Satisfying all conditions of probability distribution and visibility allows us to exclude any separable state model.



\section{Weak entanglement criterion}  
In general EPR steering scenario, there are two parities, one of which is trusted the other is untrusted. In that case, Local hidden state (LHS) model is considered \cite{EPR2} to simulate the predictions of genuine EPR steering. Fortunately, in the following, we do not need to make use of  LHS model (separable model is considered) to analyze the steering since two parities are trusted (controlled by ourselves) and the system $\mathcal{Q_A}$ and $\mathcal{Q_B}$ are genuine quantum states.  
As mentioned above, the quantum states generated by quantum mediator is $|\Psi\rangle=\text{cos}(\frac{\Delta\phi\tau}{2\hbar})|+\rangle_A|+\rangle_{B}+i\text{sin}(\frac{\Delta\phi\tau}{2\hbar})|-\rangle_A|-\rangle_{B}$ ~\cite{note}.  Here we set two measurement basis for $\mathcal{Q_A}$ are $\{|0\rangle, |1\rangle\}$ and $\{|\epsilon\rangle, |\epsilon^{\perp}\rangle\}$ respectively, we have four steered but not normalized quantum states

\begin{equation}\label{eq4}
	\begin{split}
		\tilde{\rho}^{\langle 0|_A}_{B}(Q) &=\frac{1}{2}  |\phi_+\rangle \langle \phi_+|_B,\\
		\tilde{\rho}^{\langle 1|_A}_{B}(Q)  &=\frac{1}{2} |\phi_{-}\rangle\langle \phi_{-}|_B,\\
		\tilde{\varrho}^{\langle \epsilon|_A}_{B}(Q)  &=|\alpha|^2 \text{Tr}(|\tilde{\chi}_{\epsilon}\rangle \langle \tilde{\chi}_{\epsilon}|) |\chi_{\epsilon}\rangle \langle \chi_{\epsilon}|_B,\\
		\tilde{\varrho}^{\langle \epsilon^{\perp}|_A}_{B}(Q)  &=|\beta|^2 \text{Tr}(|\tilde{\chi}_{\epsilon^{\perp}}\rangle\langle \tilde{\chi}_{\epsilon^{\perp}}|) |\chi_{\epsilon^{\perp}}\rangle\langle \chi_{\epsilon^{\perp}}|_B,\\
	\end{split}
\end{equation}
where $\alpha= \langle \epsilon|+\rangle$, $\beta=\langle \epsilon^{\perp}|+\rangle$, $|\phi_{\pm}\rangle=\text{cos}(\frac{\Delta\phi\tau}{2\hbar})|+\rangle \pm i\text{sin}(\frac{\Delta\phi\tau}{2\hbar})|-\rangle$,
$|\chi_{\epsilon}\rangle \langle \chi_{\epsilon}|=\frac{|\tilde{\chi}_{\epsilon}\rangle \langle \tilde{\chi}_{\epsilon}|}{\text{Tr}(|\tilde{\chi}_{\epsilon}\rangle \langle \tilde{\chi}_{\epsilon}|)}$, $|\chi_{\epsilon^{\perp}}\rangle\langle \chi_{\epsilon^{\perp}}|=\frac{|\tilde{\chi}_{\epsilon^{\perp}}\rangle\langle \tilde{\chi}_{\epsilon^{\perp}}|}{\text{Tr}(|\tilde{\chi}_{\epsilon^{\perp}}\rangle\langle \tilde{\chi}_{\epsilon^{\perp}}|)}$ and
$\tilde{\varrho}_{B}$ represent the unnormalized quantum states ($(Q)$ represents the quantum mediator).  Eqn. (\ref{eq4}) indicates that when we project the quantum states of $\mathcal{Q_A}$ to 
$|0\rangle, |1\rangle, |\epsilon\rangle, |\epsilon^{\perp}\rangle\}$, we get the quantum states of $\mathcal{Q_B}$ are $|\phi_+\rangle, |\phi_{-}\rangle, |\chi_{\epsilon}\rangle, |\chi_{\epsilon^{\perp}}\rangle$  with probabilities
\begin{equation}\label{probb}
	\begin{split}
&\{p^{(0,\phi_+)}, p^{(1,\phi_-)},p^{(\epsilon,\chi_{\epsilon})}, p^{(\epsilon^{\perp},\chi_{\epsilon^{\perp}})}\}  \\
& = \{\frac{1}{2},\frac{1}{2},|\alpha|^2 \text{Tr}(|\tilde{\chi}_{\epsilon}\rangle \langle \tilde{\chi}_{\epsilon}|),|\beta|^2 \text{Tr}(|\tilde{\chi}_{\epsilon^{\perp}}\rangle\langle \tilde{\chi}_{\epsilon^{\perp}}|) \},
	\end{split}
\end{equation}
  respectively. If there exists a separable model (classical mediator) can fake the results of Eqn. (\ref{probb}), then one does not
convinced that  $\mathcal{Q_A}$ can steer $\mathcal{Q_B}$'s quantum state (namely, $\mathcal{Q_A}$ and $\mathcal{Q_B}$ are unentangled). Otherwise the separable model contradicts with the quantum predictions.  However, since the precision of the measurement devices is limited \cite{projective}, we may not be able to measure the signal of weak entanglement. 
One can verify  that when the entanglement is extremely weak (i.e., $\frac{\Delta\phi\tau}{2\hbar}$  is very small), the probability $p^{(\epsilon,\chi_{\epsilon})}=|\alpha|^2 \text{Tr}(|\tilde{\chi}_{\epsilon}\rangle \langle \tilde{\chi}_{\epsilon}|)=|\alpha|^2[\text{cos}^2(\frac{\Delta\phi\tau}{2\hbar})+\text{sin}^2(\frac{\Delta\phi\tau}{2\hbar})|A^{\epsilon}_w|^2]$ will be very small so that the measurement device with resolution $\gamma\ge p^{(\epsilon,\chi_{\epsilon})}/(p^{(\epsilon,\chi_{\epsilon})}+p^{(\epsilon^{\perp},\chi_{\epsilon^{\perp}})})$ can not distinguish whether $\mathcal{Q_A}$ and $\mathcal{Q_B}$ are entangled or separable or not \cite{SI}.   
Therefore, we cannot directly determine weak entanglement in this way. Similarly, entanglement witness and other inequality methods to calculate the expectation value will also fail in this case. 

Here we note that the Eqn. \eqref{eq4} and \eqref{probb} may cover the predictions of visibility of the measurement of $\mathcal{Q_B}$ (steered state) when the heralded probability is very small, i.e. $p^{(\epsilon,\chi_{\epsilon})}\rightarrow 0$.
Here we show that the visibility of measurement of system  $\mathcal{Q_B}$ is more robust and powerful to detect weak entanglement.   Without loss generality, we define the visibility of projective measurement $\Pi_i$ is 
\begin{equation}
	V_{\Pi_i}(\varrho_B)=\text{Tr}(\Pi_i\varrho_B),
\end{equation}
 One can see that, for a pure qubit, the maximal visibility is $1$  while for a mixed state, it is impossible to obtain the visibility equals to 1. Let's set  $\{\Pi_0,\Pi_1,\Pi_2,\Pi_3\}$ are $\{|\phi_{+}\rangle \langle \phi_{+}|, |\phi_{-}\rangle \langle \phi_{-}|,|\chi_{\epsilon}\rangle  \langle \chi_{\epsilon}|,|\chi_{\epsilon^{\perp}}\rangle \langle 
 \chi_{\epsilon^{\perp}}|\}$ respectively. Now we have four sets of measurement visibility
\begin{equation} \label{vis}
\begin{split}
V_{\Pi_0}(\varrho^{\langle 0|_A}_B)&=\text{Tr}(|\phi_{+}\rangle \langle \phi_{+}|\varrho^{\langle 0|_A}_B),\\
V_{\Pi_1}(\varrho^{\langle 1|_A}_B)&=\text{Tr}(|\phi_{-}\rangle \langle \phi_{-}|\varrho^{\langle 1_A}_B ),\\
V_{\Pi_2}(\varrho^{\langle \epsilon|_A}_B)&=\text{Tr}(|\chi_{\epsilon}\rangle  \langle \chi_{\epsilon}|\varrho^{\langle \epsilon|_A}_B),\\
V_{\Pi_3}(\varrho^{\langle \epsilon^{\perp}|_A}_B)&=\text{Tr}(|\chi_{\epsilon^{\perp}}\rangle \langle 
\chi_{\epsilon^{\perp}}|\varrho^{\langle \epsilon^{\perp}|_A}_B),
\end{split}
\end{equation}
where $\varrho^{\langle 0|_A}_B,\varrho^{\langle 1|_A}_B,\varrho^{\langle \epsilon|_A}_B,\varrho^{\langle \epsilon^{\perp}|_A}_B$ are the steered and normalized quantum state of $\mathcal{Q_B}$. Obviously, the steered states of Eqn. \eqref{eq4} satisfy that all above $V$ are equal to 1. That is 
\begin{equation}\label{visq}
V^Q_{\Pi_0}(\varrho^{\langle 0|_A}_B)=V^Q_{\Pi_1}(\varrho^{\langle 1|_A}_B)=V^Q_{\Pi_2}(\varrho^{\langle \epsilon|_A}_B)=V^Q_{\Pi_3}(\varrho^{\langle \epsilon^{\perp}|_A}_B)=1,
\end{equation}
in which label $Q$ indicates quantum prediction.
As we analyzed before, in  the case of extremely weak entanglement, the results of Eqn. \eqref{probb} (probability distribution) can be simulate by a separable state $\varrho^{C}_{AB}$ \cite{SI}.  Naturally, one may wonder whether separable states $\varrho^{C}_{AB}$ can also emulate the  measurement visibility.  
Our finding is that the visibility of measurement setting $V_{\Pi_2}(\varrho^{\langle \epsilon|_A}_B)$ corresponding to the weak value amplification cannot be simulated.  Specifically,  the classical visibility of measurement $\Pi_2$ is given as \cite{SI}
\begin{widetext}

\begin{equation}\label{VIS2}
		V^{C}_{\Pi_2}(\varrho^{\langle \epsilon|_A}_B)=\text{Tr}(|\chi_{\epsilon}\rangle  \langle \chi_{\epsilon}|\varrho^{\langle \epsilon|_A}_B)=\frac{\frac{1}{2}\epsilon^2p^{(0,\phi_{+})} \langle \chi_{\epsilon}|\phi_{+}\rangle \langle \phi_{+}| \chi_{\epsilon}\rangle+\frac{1}{2}(1-\epsilon^2)p^{(1,\phi_{-})} \langle \chi_{\epsilon} |\phi_{-}\rangle \langle \phi_{-}|\chi_{\epsilon} \rangle +\frac{1}{2} p^{(\epsilon,\chi_{\epsilon})}    }{\frac{1}{2}\epsilon^2p^{(0,\phi_{+})}+\frac{1}{2}(1-\epsilon^2) p^{(1,\phi_{-})}+\frac{1}{2}p^{(\epsilon,\chi_{\epsilon})}}.
\end{equation}

\end{widetext}
When the entanglement is extremely weak, without loss generality, we set $\text{cos}(\frac{\Delta\phi\tau}{2\hbar})\approx1$, $\text{sin}(\frac{\Delta\phi\tau}{2\hbar})=\frac{\Delta\phi\tau}{2\hbar}$.  The measurement basis $\{|\epsilon\rangle, |\epsilon^{\perp}\rangle  \}$ is chosen to realize weak-value amplification (i.e. $A^{\epsilon}_w=k\frac{1}{\frac{\Delta\phi\tau}{2\hbar}}$ and $A^{\epsilon^{\perp}}_w=\frac{1}{A^{\epsilon}_w}\approx \frac{\Delta\phi\tau}{2k\hbar}\ll1 $, where $k$ is a coefficient) when $\epsilon\rightarrow \frac{1}{\sqrt{2}}$ and we have $\alpha= \langle \epsilon|+\rangle\approx0$, $\beta=\langle \epsilon^{\perp}|+\rangle\approx1$ and $\alpha k \ll1$. Upon substituting these approximations into Eqn. \eqref{VIS2} (discard the second-order small quantity $|\alpha|^2$, $(\frac{\Delta\phi\tau}{2\hbar})^2$ and set $1\pm (\frac{\Delta\phi\tau}{2\hbar})^2 \approx1$),  we obtain \cite{SI}

\begin{equation} \label{vis}
	\begin{split}
V^C_{\Pi_2}(\varrho^{\langle \epsilon|_A}_B)&\approx\frac{1}{1+k^2},\\
	\end{split}
\end{equation}
while $V^C_{\Pi_0}\approx V^C_{\Pi_1}\approx V^C_{\Pi_3}\approx 1$. One can see that this result is contradictory to the results in Eqn. \eqref{visq}. The measured visibility of  weak entanglement is all equal to $1$, however, the separable model has a $\frac{1}{1+k^2}$.  If $k=1$, we have $V^C_{\Pi_2}(\varrho^{\langle \epsilon|_A}_B)\approx\frac{1}{2}$. This is a logical contradiction of weak entanglement. It is clear that the distinguishability of measurement visibility is much greater than the probability distribution of measurement.   Therefore, the signal of weak entanglement is amplified.  Another implication of amplifying entanglement seems to be that we can reduce the experimental requirement in  tests of  gravitationally induced entanglement. Given the sensitivity of measurement device, our scheme can achieved $X=A^{\epsilon}_w$ saving for coupling strength of gravity. For example, if $A^{\epsilon}_w=10^4$, we can reduce the mass of two systems by 10 times, and shorten the coupling time by 100 times \cite{SI}.
Our methodology does not depend on a specific physical system.  Hence, different physical systems may realize amplification of gravitationally induced entanglement by this approaches.  Besides, our weak entanglement criterion remains valid at a certain degree of decoherence and the limited precision of measurement device \cite{SI}.

\section{Tests of gravitationally induced entanglement} %
As a result of technological advance in quantum manipulation of matter at larger mass scales \cite{LIGO,Delic, Tebben} and in gravitational measurements at smaller mesoscopic mass scales \cite{Westphal}, probing nonclassical nature of gravity becomes possible.  
In Ref \cite{Bose,Had},  the spin degrees of freedom of the particles are used to construct the Stern-Gerlach interferometry to test quantum gravity. Remarkably, there are many other physical realisations apart from spin degrees of freedom on probing gravitationally induced entanglement ~ \cite{CVD,Brukner,Bose2,Tanjung,Hall,Chevalier,Kamp,Kent,setup2,setup22,Bose3,Bose4,PRX,PRX2,Miao,Schut,Datta,Rijavec,Marios,CM3, Neppoleon, Bose5, Kanno,Animesh}, such as neutrino-like oscillations \cite{CVD},  optomechanics \cite{Tanjung} and atomic interferometers \cite{PRX,SI}.

Very recently , there is a promising experimental proposal that uses two-level systems coupled to a massive resonator (a harmonic oscillator) to probe gravitationally induced entanglement  \cite{MP}. Unlike Ref. \cite{PRX}, it enhance the gravitational interaction of two two-level systems by a massive particle (as a mediator), where the effective gravity-induced coupling strength is increased by a factor of $\frac{g_b}{w}$ \cite{MP}. Our criterion can also be applied to this scenario to achieve additional amplification.



\section{Discussion and conclusion}
Historically, there is a one measurement steering protocol the same as the amplification by LOCC.  In particular,  Gisin's paper in 1995, who called it ``Hidden quantum nonlocality revealed by local filters'' \cite{Gisin}.  However, this local filter is essentially a Positive Operator Valued Measures (POVM)  and needs to be performed with the help of an additional Hilbert space, such as an additional ancillary qubit. The POVM measurement will increase the experimental difficulty because it requires additional coupling to a new quantum system and measuring it. Certainly, if one do not consider the difficulty of measurement, then one may perform two types of entanglement amplification. The first type of amplification can be achieved by using the local filter method, and the second type of amplification is achieved by the way we propose in the paper. 

From a fundamental perspective, our work combines weak value theory and quantum correlation theory for the first time. We show that weak-valued amplification in the two-setting protocol is impossible to be simulated classically. Our results also support the fact that weak values are quantum, whereas in the past it was controversial whether weak values were quantum or not~\cite{weak1,weak2,weak3,weak4, weak} .


Our scheme is applicable to any weakly entangled pure state, while allowing for the presence of partial decoherence and noise. It be expected to significantly reduce the requirements for experiments, allowing for the test of gravitationally induced entanglement in the near future. From a more general point of view, our results is a general weak entanglement criterion. We reveal how the hidden weakly entangled information is re-presented as it is. Compared to the previous protocols, our approach allows us to observe entangled signals that were previously impossible to observe. As an outlook, we expect that the criterion can be extended to the more general mixed states, which may make it more possible to detect the entanglement of macroscopic objects.

\begin{acknowledgments}
TF thanks  the support of the Fulan Scholarship at Sun Yat-sen University and the National Natural
Science Foundation of China Grants No. 12147107. VV' s research is supported by the National Research Foundation and the Ministry of Education in Singapore and administered by the Centre for Quantum Technologies, National University of Singapore. This publication was made possible through the support of the ID 61466 grant from the John Templeton Foundation, as part of the The Quantum Information Structure of Spacetime (QISS) Project (qiss.fr). The opinions expressed in this publication are those of the authors and do not necessarily reflect the views of the John Templeton Foundation. 
\end{acknowledgments}


\section*{Appendix A : separable state model for classical mediator}

\subsection{Separable state model} Here we analyze the model of separable states for $\mathcal{Q_A}$ and $\mathcal{Q_B}$  (induced by a classical mediator) that can simulate the results of quantum entanglement. 
As we mentioned before, the quantum states generated by classical mediator is $\varrho_{AB}=\Lambda^C(\varrho_{A}\otimes\varrho_{B})= \sum_i p_i\varrho^i_A\otimes \varrho^i_B$, we have

\begin{equation}
	\begin{split}
		\tilde{\varrho}^{\langle 0|_A}_{B} (C) &=\sum_i p_i \text{Tr}( |0\rangle \langle 0| \varrho^i_A )\varrho^i_B,\\
		\tilde{\varrho}^{\langle 1|_A}_{B} (C) &=\sum_i p_i \text{Tr}( |1\rangle \langle 1| \varrho^i_A ) \varrho^i_B,\\
		\tilde{\varrho}^{\langle \epsilon|_A}_{B} (C) &=\sum_i p_i \text{Tr}( |\epsilon\rangle \langle \epsilon| \varrho^i_A )\varrho^i_B,\\
		\tilde{\varrho}^{\langle \epsilon^{\perp}|_A}_{B} (C) &=\sum_i p_i \text{Tr}(|\epsilon^{\perp}\rangle \langle \epsilon^{\perp}| \varrho^i_A)\varrho^i_B,\\
	\end{split}
\end{equation}
where  $\tilde{\varrho}_{B} (C)$ is the unnormalized quantum state with the classical mediator. If a calssical mediator can simulate all the results of a quantum mediator,  it must satisfy $\tilde{\varrho}^{\langle0|}_{B} (Q) =\tilde{\varrho}^{\langle 0|}_{B} (C)$,$\tilde{\varrho}^{\langle1|}_{B} (Q) =\tilde{\varrho}^{\langle 1|}_{B} (C)$, $\tilde{\varrho}^{\langle \epsilon|}_{B} (Q) =\tilde{\varrho}^{\langle \epsilon|}_{B} (C)$ and $\tilde{\varrho}^{\langle \epsilon^{\perp}|}_{B} (Q)=\tilde{\varrho}^{\langle \epsilon^{\perp}|}_{B} (C)$. So we have 
\begin{equation}
	\begin{split}	
	&	p^{(0,\chi_{\phi_+})} |0\rangle \langle 0| _A\otimes |\phi_+\rangle \langle \phi_+|_B \\
		&=\sum_i p_i \text{Tr}( |0\rangle \langle0| \varrho^i_A )\text{Tr}(|\phi_+\rangle \langle\phi_+|\varrho^i_B) \varrho^i_{A}\otimes \varrho^i_{B},\\
	&	p^{(1,\chi_{\phi_-})} |1\rangle \langle 1| _A\otimes |\phi_-\rangle \langle \phi_-|_B \\
		&=\sum_i p_i \text{Tr}( |1\rangle \langle1| \varrho^i_A )\text{Tr}(|\phi_-\rangle \langle\phi_-|\varrho^i_B) \varrho^i_{A}\otimes \varrho^i_{B} ,\\
	&	p^{(\epsilon,\chi_{\epsilon})}|\epsilon\rangle \langle \epsilon| _A\otimes |\chi_{\epsilon}\rangle \langle \chi_{\epsilon}|_B \\
		&=\sum_i p_i \text{Tr}( |\epsilon\rangle \langle \epsilon| \varrho^i_A )\text{Tr}(|\chi_{\epsilon}\rangle \langle \chi_{\epsilon}|\varrho^i_B) \varrho^i_{A}\otimes \varrho^i_{B},\\
	&	p^{(\epsilon^{\perp},\chi_{\epsilon^{\perp}})} |\epsilon^{\perp}\rangle \langle \epsilon^{\perp}| _A\otimes |\chi_{\epsilon^{\perp}}\rangle \langle \chi_{\epsilon^{\perp}}|_B \\
		&=\sum_i p_i \text{Tr}( |\epsilon^{\perp}\rangle \langle \epsilon^{\perp}| \varrho^i_A ) \text{Tr}(|\chi_{\epsilon}\rangle \langle \chi_{\epsilon^{\perp}}|\varrho^i_B)\varrho^i_{A}\otimes \varrho^i_{B},
	\end{split}\label{qc}
\end{equation}
where 
\begin{equation}\label{probb_C}
	\begin{split}
	\{p^{(0,\phi_+)}, p^{(1,\phi_-)},p^{(\epsilon,\chi_{\epsilon})}, p^{(\epsilon^{\perp},\chi_{\epsilon^{\perp}})}\}  \\
	 = \{\frac{1}{2},\frac{1}{2},\alpha^2 \text{Tr}(|\tilde{\chi}_{\epsilon}\rangle \langle \tilde{\chi}_{\epsilon}|),\beta^2 \text{Tr}(|\tilde{\chi}_{\epsilon^{\perp}}\rangle\langle \tilde{\chi}_{\epsilon^{\perp}}|) \}.
 \end{split}
\end{equation}

It is well-known that a pure state cannot be obtained by a convex sum of other different states, namely, a density matrix of pure state can only be expanded by itself. Let us look at Eq. (\ref{qc}), because the left-hand side is proportional to a pure state, without loss of generality, one has

\begin{widetext}
\begin{equation}
	\begin{split}
		p^{(0,\phi_+)}=p_j\text{Tr}( |0\rangle \langle 0| \varrho^j_A ) \text{Tr}( |\phi_+\rangle \langle \phi_+| \varrho^j_B ) \quad &\text{and} \quad  \varrho_A^j\otimes \varrho_B^j=|0\rangle \langle 0|_A \otimes |\phi_+\rangle \langle \phi_+|_B, \\	
		p^{(1,\phi_-)}=p_k\text{Tr}( |1\rangle \langle 1| \varrho^k_A ) \text{Tr}( |\phi_-\rangle \langle \phi_-| \varrho^k_B ) \quad &\text{and} \quad  \varrho_A^k\otimes \varrho_B^k=|1\rangle \langle 1|_A \otimes |\phi_-\rangle \langle \phi_-|_B, \\	
		p^{(\epsilon,\chi_{\epsilon})}=p_m\text{Tr}( |\epsilon\rangle \langle \epsilon| \varrho^m_A ) \text{Tr}( |\chi_{\epsilon}\rangle \langle \chi_{\epsilon}| \varrho^m_B ) \quad &\text{and} \quad  \varrho_A^m\otimes \varrho_B^m=|\epsilon\rangle \langle \epsilon|_A\otimes |\chi_{\epsilon}\rangle \langle \chi_{\epsilon}|_B,\\
		p^{(\epsilon^{\perp},\chi_{\epsilon^{\perp}})} =p_n\text{Tr}( |\epsilon^{\perp}\rangle \langle \epsilon^{\perp}|\varrho^n_A ) \text{Tr}(|\chi_{\epsilon^{\perp}}\rangle \langle \chi_{\epsilon^{\perp}} |\varrho^n_{B})\quad &\text{and} \quad \varrho_A^n\otimes\varrho_B^n=|\epsilon^{\perp}\rangle \langle \epsilon^{\perp}|_A\otimes|\chi_{\epsilon^{\perp}}\rangle\langle \chi_{\epsilon^{\perp}}|_B.
	\end{split}\label{qc2}
\end{equation}
\end{widetext}
Naturally, one may consider mixing these four pure states with corresponding probabilities to construct a separable model to simulate the prediciton of quantum mediator. Since two basis $a=\{0,1\}$ and $b=\{\epsilon,\epsilon^{\perp}\}$ are randomly selected (with probability $\frac{1}{2}$). Therefore, the separable state induced by classical mediator can be written as 
\begin{widetext}
\begin{equation}
	\begin{split} \label{C}
		\varrho^C_{AB}=\Lambda^C(\varrho_{A}\otimes \varrho_B)=
		\frac{1}{2} [ p^{(0,\phi_+)}|0\rangle \langle 0|_A \otimes |\phi_+\rangle \langle \phi_+|_B+p^{(1,\phi_-)}|1\rangle \langle 1|_A \otimes |\phi_-\rangle \langle \phi_-|_B]+ \\
		\frac{1}{2} [	p^{(\epsilon,\chi_{\epsilon})} |\epsilon\rangle  \langle \epsilon|_A \otimes |\chi_{\epsilon}\rangle \langle \chi_{\epsilon}|_B+p^{(\epsilon^{\perp},\chi_{\epsilon^{\perp}})}  |\epsilon^{\perp}\rangle  \langle \epsilon^{\perp}|_A\otimes |\chi_{\epsilon^{\perp}}\rangle\langle \chi_{\epsilon^{\perp}}|_B].
	\end{split}
\end{equation}
\end{widetext}
It should be noted that if our test only using a single basis $b=\{\epsilon,\epsilon^{\perp}\}$, it is easy to find a separable state to simulate the results of quantum mediator (one can verify it). Similar to quantum steering scenario, two or more than two basis are considered,
in theory, there is no classcal quantum mediator can simulate it ( probability distribution).  However,  ideal projective measurements can not be implemented in experiments since they need infinite resource costs \cite{projective}. That is, the measurement device has limited resolution.  One can verify  that when the entanglement is extremely weak (i.e., $\frac{\Delta\phi\tau}{2\hbar}$  is very small), the probability $p^{(\epsilon,\chi_{\epsilon})}=\alpha^2 \text{Tr}(|\tilde{\chi}_{\epsilon}\rangle \langle \tilde{\chi}_{\epsilon}|)=\alpha^2[\text{cos}^2(\frac{\Delta\phi\tau}{2\hbar})+\text{sin}^2(\frac{\Delta\phi\tau}{2\hbar})|A^{\epsilon}_w|^2]$ will be very small so that the measurement device with resolution $\eta\ge p^{(\epsilon,\chi_{\epsilon})}/(p^{(\epsilon,\chi_{\epsilon})}+p^{(\epsilon^{\perp},\chi_{\epsilon^{\perp}})})$ can not distinguish  whether $\mathcal{Q_A}$ and $\mathcal{Q_B}$ are entangled or separable or not. 


\emph{Example.} Suppose $\frac{\Delta\phi\tau}{2\hbar}$ is very small and $A^{\epsilon}_w$  is very large, we approximate $\text{cos}(\frac{\Delta\phi\tau}{2\hbar})\approx 1$ and $\text{sin}(\frac{\Delta\phi\tau}{2\hbar}) \approx  \frac{\Delta\phi\tau}{2\hbar}$. So we have $| \tilde{\chi}_{\epsilon^{\perp}}\rangle \approx |+\rangle$ and $|\tilde{\chi}_{\epsilon}\rangle =	 |+\rangle+i\frac{\Delta\phi\tau}{2\hbar}A^{\epsilon}_w  |-\rangle$. Here we set  $\frac{\Delta\phi\tau}{2\hbar}A^{\epsilon}_w=1$, the quantum state $|\tilde{\chi}_{\epsilon}\rangle$ becomes $|+\rangle+i|-\rangle$.  So we have

\begin{equation}\label{exa}
	\{p^{(0,\phi_+)}, p^{(1,\phi_-)},p^{(\epsilon,\chi_{\epsilon})}, p^{(\epsilon^{\perp},\chi_{\epsilon^{\perp}})}\}   = \{\frac{1}{2},\frac{1}{2},2\alpha^2 ,\beta^2 \}.
\end{equation}
Recall $\alpha= \langle \epsilon|+\rangle$, $\beta=\langle \epsilon^{\perp}|+\rangle$,  $|\epsilon\rangle=\epsilon|0\rangle-\sqrt{1-\epsilon^2}|1\rangle$ and $|\epsilon^{\perp}\rangle=\sqrt{1-\epsilon^2}|0\rangle+\epsilon|1\rangle$. If we want to get a big weak value $A^{\epsilon}_w=\frac{\langle \epsilon|Z|+\rangle}{\langle \epsilon|+\rangle}=\frac{\epsilon+\sqrt{1-\epsilon^2}}{\epsilon-\sqrt{1-\epsilon^2}}$, $\epsilon$ should close to $\frac{1}{\sqrt{2}}$.  So $\alpha\approx \frac{\sqrt{2}}{A^{\epsilon}_w}=\sqrt{2}\frac{\Delta\phi\tau}{2\hbar}$ and $\beta \approx1$.  Therefore, Eqn. (\ref{exa}) becomes
\begin{equation}
	\{p^{(0,+)}, p^{(1,+)},p^{(-,\chi_{\epsilon})}, p^{(+,+)}\}   = \{\frac{1}{2},\frac{1}{2},4(\frac{\Delta\phi\tau}{2\hbar})^2 ,1\}.
\end{equation}
It is easy to verify that the above results can be simulated by a separable state $|+\rangle_A \otimes |+\rangle_B$ (a more accurate model should be in the form of $\varrho^C_{AB}$) if $4(\frac{\Delta\phi\tau}{2\hbar})^2$ is small which may be masked by the noise of measurement device.\\


\begin{widetext}

\subsection{Visibility of measurement for separable model} Before calculate the visibility of measurement, we need to find the reduced density matrix $\varrho^{\langle 0|_A}_B,\varrho^{\langle 1|_A}_B,\varrho^{\langle \epsilon|_A}_B$ and $ \varrho^{\langle \epsilon^{\perp}|_A}_B$ for $\varrho^{C}_{AB}$, which are as follow

\begin{equation}
	\begin{split}
		\varrho^{\langle 0|_A}_B&=\frac{\text{Tr}_A(|0\rangle \langle 0|_A\otimes I_B \varrho^{C}_{AB} )}{\text{Tr}(|0\rangle \langle 0|_A\otimes I_B \varrho^{C}_{AB} )}=\frac{\frac{1}{2}p^{(0,\phi_{+})}|\phi_{+}\rangle \langle \phi_{+}|+\frac{1}{2}\epsilon^2 p^{(\epsilon,\chi_{\epsilon})} |\chi_{\epsilon}\rangle \langle \chi_{\epsilon}| + \frac{1}{2}(1-\epsilon^2)p^{(\epsilon^{\perp},\chi_{\epsilon^{\perp}})}|\chi_{\epsilon^{\perp}}\rangle \langle \chi_{\epsilon^{\perp}}|    }{\frac{1}{2}p^{(0,\phi_{+})}+\frac{1}{2}\epsilon^2 p^{(\epsilon,\chi_{\epsilon})}+\frac{1}{2}(1-\epsilon^2)p^{(\epsilon^{\perp},\chi_{\epsilon^{\perp}})}};\\ 
		\varrho^{\langle 1|_A}_B&=\frac{\text{Tr}_A(|1\rangle \langle 1|_A\otimes I_B \varrho^{C}_{AB} )}{\text{Tr}(|1\rangle \langle 1|_A\otimes I_B \varrho^{C}_{AB} )}=\frac{\frac{1}{2}p^{(1,\phi_{-})}|\phi_{-}\rangle \langle \phi_{-}|+\frac{1}{2}(1-\epsilon^2) p^{(\epsilon,\chi_{\epsilon})} |\chi_{\epsilon}\rangle \langle \chi_{\epsilon}| + \frac{1}{2}\epsilon^2 p^{(\epsilon^{\perp},\chi_{\epsilon^{\perp}})}|\chi_{\epsilon^{\perp}}\rangle \langle \chi_{\epsilon^{\perp}}|    }{\frac{1}{2}p^{(1,\phi_{-})}+\frac{1}{2}(1-\epsilon^2) p^{(\epsilon,\chi_{\epsilon})}+\frac{1}{2}\epsilon^2p^{(\epsilon^{\perp},\chi_{\epsilon^{\perp}})}};\\
		\varrho^{\langle \epsilon|_A}_B&=\frac{\text{Tr}_A(|\epsilon\rangle \langle \epsilon|_A\otimes I_B \varrho^{C}_{AB} )}{\text{Tr}(|\epsilon\rangle \langle \epsilon|_A\otimes I_B \varrho^{C}_{AB} )}=\frac{\frac{1}{2}\epsilon^2p^{(0,\phi_{+})}|\phi_{+}\rangle \langle \phi_{+}|+\frac{1}{2}(1-\epsilon^2)p^{(1,\phi_{-})}|\phi_{-}\rangle \langle \phi_{-}|+\frac{1}{2} p^{(\epsilon,\chi_{\epsilon})} |\chi_{\epsilon}\rangle \langle \chi_{\epsilon}|    }{\frac{1}{2}\epsilon^2p^{(0,\phi_{+})}+\frac{1}{2}(1-\epsilon^2) p^{(1,\phi_{-})}+\frac{1}{2}p^{(\epsilon,\chi_{\epsilon})}};\\
		\varrho^{\langle \epsilon^{\perp}|_A}_B&=\frac{\text{Tr}_A(|\epsilon^{\perp}\rangle \langle \epsilon^{\perp}|_A\otimes I_B \varrho^{C}_{AB} )}{\text{Tr}(|\epsilon^{\perp}\rangle \langle \epsilon^{\perp}|_A\otimes I_B \varrho^{C}_{AB} )}=\frac{\frac{1}{2}(1-\epsilon^2)p^{(0,\phi_{+})}|\phi_{+}\rangle \langle \phi_{+}|+\frac{1}{2}\epsilon^2p^{(1,\phi_{-})}|\phi_{-}\rangle \langle \phi_{-}|+\frac{1}{2} p^{(\epsilon^{\perp},\chi_{\epsilon^{\perp}})} |\chi_{\epsilon^{\perp}}\rangle \langle \chi_{\epsilon^{\perp}}|    }{\frac{1}{2}(1-\epsilon^2)p^{(0,\phi_{+})}+\frac{1}{2}\epsilon^2 p^{(\epsilon,\chi_{\epsilon})}+\frac{1}{2}p^{(\epsilon^{\perp},\chi_{\epsilon^{\perp}})}}.
	\end{split}
\end{equation}

Now substitute them into the Eqn. \eqref{vis}, we get the visibility of measurement of $\varrho^C_{AB}$:

\begin{equation} \label{vis_C}
	\begin{split}
		V^C_{\Pi_0}(\varrho^{\langle 0|_A}_B)&=\text{Tr}(|\phi_{+}\rangle \langle \phi_{+}|\varrho^{\langle 0|_A}_B)
		=\frac{\frac{1}{2}p^{(0,\phi_{+})}+\frac{1}{2}\epsilon^2 p^{(\epsilon,\chi_{\epsilon})} \langle \phi_{+}|\chi_{\epsilon}\rangle \langle \chi_{\epsilon}| \phi_{+}\rangle + \frac{1}{2}(1-\epsilon^2)p^{(\epsilon^{\perp},\chi_{\epsilon^{\perp}})} \langle \phi_{+}|\chi_{\epsilon^{\perp}}\rangle \langle \chi_{\epsilon^{\perp}}|  \phi_{+} \rangle }{\frac{1}{2}p^{(0,\phi_{+})}+\frac{1}{2}\epsilon^2 p^{(\epsilon,\chi_{\epsilon})}+\frac{1}{2}(1-\epsilon^2)p^{(\epsilon^{\perp},\chi_{\epsilon^{\perp}})}};\\ 
		V^C_{\Pi_1}(\varrho^{\langle 1|_A}_B)&=\text{Tr}(|\phi_{-}\rangle \langle \phi_{-}|\varrho^{\langle 1_A}_B )=\frac{\frac{1}{2}p^{(1,\phi_{-})}+\frac{1}{2}(1-\epsilon^2) p^{(\epsilon,\chi_{\epsilon})} \langle \phi_{-}|\chi_{\epsilon}\rangle \langle \chi_{\epsilon}|\phi_{-}\rangle + \frac{1}{2}\epsilon^2 p^{(\epsilon^{\perp},\chi_{\epsilon^{\perp}})} \langle \phi_{-}|\chi_{\epsilon^{\perp}}\rangle \langle \chi_{\epsilon^{\perp}}| \phi_{-}\rangle   }{\frac{1}{2}p^{(1,\phi_{-})}+\frac{1}{2}(1-\epsilon^2) p^{(\epsilon,\chi_{\epsilon})}+\frac{1}{2}\epsilon^2p^{(\epsilon^{\perp},\chi_{\epsilon^{\perp}})}};\\
		V^{C}_{\Pi_2}(\varrho^{\langle \epsilon|_A}_B)&=\text{Tr}(|\chi_{\epsilon}\rangle  \langle \chi_{\epsilon}|\varrho^{\langle \epsilon|_A}_B)=\frac{\frac{1}{2}\epsilon^2p^{(0,\phi_{+})} \langle \chi_{\epsilon}|\phi_{+}\rangle \langle \phi_{+}| \chi_{\epsilon}\rangle+\frac{1}{2}(1-\epsilon^2)p^{(1,\phi_{-})} \langle \chi_{\epsilon} |\phi_{-}\rangle \langle \phi_{-}|\chi_{\epsilon} \rangle +\frac{1}{2} p^{(\epsilon,\chi_{\epsilon})}    }{\frac{1}{2}\epsilon^2p^{(0,\phi_{+})}+\frac{1}{2}(1-\epsilon^2) p^{(1,\phi_{-})}+\frac{1}{2}p^{(\epsilon,\chi_{\epsilon})}};\\
		V^{C}_{\Pi_3}(\varrho^{\langle \epsilon^{\perp}|_A}_B)&=\text{Tr}(|\chi_{\epsilon^{\perp}}\rangle \langle 
		\chi_{\epsilon^{\perp}}|\varrho^{\langle \epsilon^{\perp}|_A}_B)=\frac{\frac{1}{2}(1-\epsilon^2)p^{(0,\phi_{+})} \langle \chi_{\epsilon^{\perp}} |\phi_{+}\rangle \langle \phi_{+}| \chi_{\epsilon^{\perp}}\rangle+\frac{1}{2}\epsilon^2p^{(1,\phi_{-})}  \langle  \chi_{\epsilon^{\perp}}|\phi_{-}\rangle \langle \phi_{-}|\chi_{\epsilon^{\perp}}\rangle+\frac{1}{2} p^{(\epsilon^{\perp},\chi_{\epsilon^{\perp}})}    }{\frac{1}{2}(1-\epsilon^2)p^{(0,\phi_{+})}+\frac{1}{2}\epsilon^2 p^{(\epsilon,\chi_{\epsilon})}+\frac{1}{2}p^{(\epsilon^{\perp},\chi_{\epsilon^{\perp}})}},
	\end{split}
\end{equation}

where $|\phi_{\pm}\rangle=\text{cos}(\frac{\Delta\phi\tau}{2\hbar})|+\rangle \pm i\text{sin}(\frac{\Delta\phi\tau}{2\hbar})|-\rangle$, 
$	|\chi_{\epsilon}\rangle=	\frac{\text{cos}(\frac{\Delta\phi\tau}{2\hbar})  |+\rangle_B+i\text{sin}(\frac{\Delta\phi\tau}{2\hbar})A^{\epsilon}_w  |-\rangle}{[\text{cos}^2(\frac{\Delta\phi\tau}{2\hbar})+\text{sin}^2(\frac{\Delta\phi\tau}{2\hbar})|A^{\epsilon}_w|^2]^{\frac{1}{2}}},$,
and
$	|\chi_{\epsilon^{\perp}}\rangle=\frac{	\text{cos}(\frac{\Delta\phi\tau}{2\hbar})  |+\rangle+i\text{sin}(\frac{\Delta\phi\tau}{2\hbar})A^{\epsilon^{\perp}}_w  |-\rangle}{[\text{cos}^2(\frac{\Delta\phi\tau}{2\hbar})+\text{sin}^2(\frac{\Delta\phi\tau}{2\hbar})|A^{\epsilon^{\perp}}_w|^2]^{\frac{1}{2}}}$. 

Recall that $\{p^{(0,\phi_+)}, p^{(1,\phi_-)},p^{(\epsilon,\chi_{\epsilon})}, p^{(\epsilon^{\perp},\chi_{\epsilon^{\perp}})}\}   = \{\frac{1}{2},\frac{1}{2},\alpha^2 \text{Tr}(|\tilde{\chi}_{\epsilon}\rangle \langle \tilde{\chi}_{\epsilon}|),\beta^2 \text{Tr}(|\tilde{\chi}_{\epsilon^{\perp}}\rangle\langle \tilde{\chi}_{\epsilon^{\perp}}|) \}$, $\text{Tr}(|\tilde{\chi}_{\epsilon}\rangle \langle \tilde{\chi}_{\epsilon}|)=\text{cos}^2(\frac{\Delta\phi\tau}{2\hbar})+\text{sin}^2(\frac{\Delta\phi\tau}{2\hbar})|A^{\epsilon}_w|^2$, $\text{Tr}(|\tilde{\chi}_{\epsilon^{\perp}}\rangle \langle \tilde{\chi}_{\epsilon^{\perp}}|)=\text{cos}^2(\frac{\Delta\phi\tau}{2\hbar})+\text{sin}^2(\frac{\Delta\phi\tau}{2\hbar})|A^{\epsilon^{\perp}}_w|^2$, 
$|\langle \phi_{+}| \chi_{\epsilon}\rangle|^2=\frac{|\text{cos}^2(\frac{\Delta\phi\tau}{2\hbar})+\text{sin}^2(\frac{\Delta\phi\tau}{2\hbar})A^{\epsilon}_w|^2}{[\text{cos}^2(\frac{\Delta\phi\tau}{2\hbar})+\text{sin}^2(\frac{\Delta\phi\tau}{2\hbar})|A^{\epsilon}_w|^2]}$, 
$|\langle \phi_{+}| \chi_{\epsilon^{\perp}}\rangle|^2=\frac{|\text{cos}^2(\frac{\Delta\phi\tau}{2\hbar})+\text{sin}^2(\frac{\Delta\phi\tau}{2\hbar})A^{\epsilon^{\perp}}_w|^2}{[\text{cos}^2(\frac{\Delta\phi\tau}{2\hbar})+\text{sin}^2(\frac{\Delta\phi\tau}{2\hbar})|A^{\epsilon^{\perp}}_w|^2]}$, 
$|\langle \phi_{-}| \chi_{\epsilon}\rangle|^2=\frac{|\text{cos}^2(\frac{\Delta\phi\tau}{2\hbar})-\text{sin}^2(\frac{\Delta\phi\tau}{2\hbar})A^{\epsilon}_w|^2}{[\text{cos}^2(\frac{\Delta\phi\tau}{2\hbar})+\text{sin}^2(\frac{\Delta\phi\tau}{2\hbar})|A^{\epsilon}_w|^2]}$, 
$|\langle \phi_{-}| \chi_{\epsilon^{\perp}}\rangle|^2=\frac{|\text{cos}^2(\frac{\Delta\phi\tau}{2\hbar})-\text{sin}^2(\frac{\Delta\phi\tau}{2\hbar})A^{\epsilon^{\perp}}_w|^2}{[\text{cos}^2(\frac{\Delta\phi\tau}{2\hbar})+\text{sin}^2(\frac{\Delta\phi\tau}{2\hbar})|A^{\epsilon^{\perp}}_w|^2]}$,  one has

\begin{equation} \label{vis_CCC}
	\begin{split}
		V^C_{\Pi_0}(\varrho^{\langle 0|_A}_B)&=
		\frac{\frac{1}{4}+\frac{1}{2}\epsilon^2 \alpha ^2|\text{cos}^2(\frac{\Delta\phi\tau}{2\hbar})+\text{sin}^2(\frac{\Delta\phi\tau}{2\hbar})A^{\epsilon}_w|^2 + \frac{1}{2}(1-\epsilon^2)\beta^2 |\text{cos}^2(\frac{\Delta\phi\tau}{2\hbar})+\text{sin}^2(\frac{\Delta\phi\tau}{2\hbar})A^{\epsilon^{\perp}}_w|^2 }{\frac{1}{4}+\frac{1}{2}\epsilon^2 \alpha ^2 [\text{cos}^2(\frac{\Delta\phi\tau}{2\hbar})+\text{sin}^2(\frac{\Delta\phi\tau}{2\hbar})|A^{\epsilon}_w|^2]
			+\frac{1}{2}(1-\epsilon^2)\beta^2[\text{cos}^2(\frac{\Delta\phi\tau}{2\hbar})+\text{sin}^2(\frac{\Delta\phi\tau}{2\hbar})|A^{\epsilon^{\perp}}_w|^2]};\\ 
		V^C_{\Pi_1}(\varrho^{\langle 1|_A}_B)&=\frac{\frac{1}{4}+\frac{1}{2}(1-\epsilon^2) \alpha^2 |\text{cos}^2(\frac{\Delta\phi\tau}{2\hbar})-\text{sin}^2(\frac{\Delta\phi\tau}{2\hbar})A^{\epsilon}_w|^2 + \frac{1}{2}\epsilon^2\beta^2|\text{cos}^2(\frac{\Delta\phi\tau}{2\hbar})-\text{sin}^2(\frac{\Delta\phi\tau}{2\hbar})A^{\epsilon^{\perp}}_w|^2 }{\frac{1}{4}+\frac{1}{2}(1-\epsilon^2) \alpha ^2 [\text{cos}^2(\frac{\Delta\phi\tau}{2\hbar})+\text{sin}^2(\frac{\Delta\phi\tau}{2\hbar})|A^{\epsilon}_w|^2]
			+\frac{1}{2}\epsilon^2\beta^2[\text{cos}^2(\frac{\Delta\phi\tau}{2\hbar})+\text{sin}^2(\frac{\Delta\phi\tau}{2\hbar})|A^{\epsilon^{\perp}}_w|^2]};\\
		V^C_{\Pi_2}(\varrho^{\langle \epsilon|_A}_B)&=\frac{\frac{1}{4}\epsilon^2 \frac{|\text{cos}^2(\frac{\Delta\phi\tau}{2\hbar})+\text{sin}^2(\frac{\Delta\phi\tau}{2\hbar})A^{\epsilon}_w|^2}{[\text{cos}^2(\frac{\Delta\phi\tau}{2\hbar})+\text{sin}^2(\frac{\Delta\phi\tau}{2\hbar})|A^{\epsilon}_w|^2]}+\frac{1}{4}(1-\epsilon^2) \frac{|\text{cos}^2(\frac{\Delta\phi\tau}{2\hbar})-\text{sin}^2(\frac{\Delta\phi\tau}{2\hbar})A^{\epsilon}_w|^2}{[\text{cos}^2(\frac{\Delta\phi\tau}{2\hbar})+\text{sin}^2(\frac{\Delta\phi\tau}{2\hbar})|A^{\epsilon}_w|^2]} +\frac{1}{2}\alpha^2 [\text{cos}^2(\frac{\Delta\phi\tau}{2\hbar})+\text{sin}^2(\frac{\Delta\phi\tau}{2\hbar})|A^{\epsilon}_w|^2] }{\frac{1}{4}\epsilon^2+\frac{1}{4}(1-\epsilon^2)+\frac{1}{2}\alpha^2 [\text{cos}^2(\frac{\Delta\phi\tau}{2\hbar})+\text{sin}^2(\frac{\Delta\phi\tau}{2\hbar})|A^{\epsilon}_w|^2]};\\
		V^C_{\Pi_3}(\varrho^{\langle \epsilon^{\perp}|_A}_B)&=\frac{\frac{1}{4}(1-\epsilon^2)\frac{|\text{cos}^2(\frac{\Delta\phi\tau}{2\hbar})+\text{sin}^2(\frac{\Delta\phi\tau}{2\hbar})A^{\epsilon^{\perp}}_w|^2}{[\text{cos}^2(\frac{\Delta\phi\tau}{2\hbar})+\text{sin}^2(\frac{\Delta\phi\tau}{2\hbar})|A^{\epsilon^{\perp}}_w|^2]}+\frac{1}{4}\epsilon^2  \frac{|\text{cos}^2(\frac{\Delta\phi\tau}{2\hbar})-\text{sin}^2(\frac{\Delta\phi\tau}{2\hbar})A^{\epsilon^{\perp}}_w|^2}{[\text{cos}^2(\frac{\Delta\phi\tau}{2\hbar})+\text{sin}^2(\frac{\Delta\phi\tau}{2\hbar})|A^{\epsilon^{\perp}}_w|^2]}+\frac{1}{2} \beta^2 \text{cos}^2(\frac{\Delta\phi\tau}{2\hbar})+\text{sin}^2(\frac{\Delta\phi\tau}{2\hbar})|A^{\epsilon^{\perp}}_w|^2   }{\frac{1}{4}(1-\epsilon^2)+\frac{1}{4}\epsilon^2 +\frac{1}{2}\beta^2 [\text{cos}^2(\frac{\Delta\phi\tau}{2\hbar})+\text{sin}^2(\frac{\Delta\phi\tau}{2\hbar})|A^{\epsilon^{\perp}}_w|^2]}.
	\end{split}
\end{equation}
\end{widetext}

\subsection{Examples} When the entanglement is extremely weak, $\text{cos}(\frac{\Delta\phi\tau}{2\hbar})\approx1$, $\text{sin}(\frac{\Delta\phi\tau}{2\hbar})=\frac{\Delta\phi\tau}{2\hbar}\sim 0$.  The measurement basis $\{|\epsilon\rangle, |\epsilon^{\perp}\rangle  \}$ can realize weak-value amplification (i.e. $A^{\epsilon}_w\approx \frac{1}{\frac{\Delta\phi\tau}{2\hbar}}$ and $A^{\epsilon^{\perp}}_w=\frac{1}{A^{\epsilon}_w}\approx \frac{\Delta\phi\tau}{2\hbar}\ll1$) when $\epsilon\rightarrow \frac{1}{\sqrt{2}}$ and we have $\alpha= \langle \epsilon|+\rangle\approx0$, $\beta=\langle \epsilon^{\perp}|+\rangle\approx1$. Upon substituting these approximations into Eqn. \eqref{vis_CCC} (discard the second-order small quantity $|\alpha|^2$, $(\frac{\Delta\phi\tau}{2\hbar})^2$ and set $1\pm (\frac{\Delta\phi\tau}{2\hbar})^2 \approx1$),  we obtain

\begin{equation} \label{visc}
	\begin{split}
		V^C_{\Pi_0}(\varrho^{\langle 0|_A}_B)&\approx 1;  \\
		V^C_{\Pi_1}(\varrho^{\langle 1|_A}_B)&\approx1 ;  \\
		V^C_{\Pi_2}(\varrho^{\langle \epsilon|_A}_B)&\approx\frac{1}{2} ;\\
		V^C_{\Pi_3}(\varrho^{\langle \epsilon^{\perp}|_A}_B)&\approx1.
	\end{split}
\end{equation}
One can see that this result is contradictory to the results analyzed in our main text. The measured visibility of  weak entanglement is all equal to $1$, however, the separable model has a $\frac{1}{2}$.  This is a logical contradiction.

The more general case is that  we set $A^{\epsilon}_w= k \frac{1}{\frac{\Delta\phi\tau}{2\hbar}}$, where is the coefficient. We also assure 
$\text{cos}(\frac{\Delta\phi\tau}{2\hbar})\approx1$, $\text{sin}(\frac{\Delta\phi\tau}{2\hbar})=\frac{\Delta\phi\tau}{2\hbar}$, $A^{\epsilon^{\perp}}_w=\frac{1}{A^{\epsilon}_w}\approx \frac{\Delta\phi\tau}{2k\hbar}\ll 1$ and $\alpha k\ll1$. One has 

\begin{equation} \label{viscck}
	\begin{split}
		V^C_{\Pi_0}(\varrho^{\langle 0|_A}_B)&\approx 1;  \\
		V^C_{\Pi_1}(\varrho^{\langle 1|_A}_B)&\approx1 ;  \\
		V^C_{\Pi_2}(\varrho^{\langle \epsilon|_A}_B)&\approx\frac{1}{1+k^2} ;\\
		V^C_{\Pi_3}(\varrho^{\langle \epsilon^{\perp}|_A}_B)&\approx1.
	\end{split}
\end{equation}

\section*{Appendix B: Experimental consideration}
We have shown that steering scenarios with weak value amplification are not possible to be simulated classically. In the following, we focus on the analysis of the steered state for weak-value amplification, so as to consider the feasibility of the experiment for detecting weak-gravitationally-induced entanglement. Let set  $m_1=m_2=m$ and suppose that the coupling strength of gravity is weak. After the weak gravity interaction and projective measurement ($\langle \epsilon|$) on $\mathcal{Q_A}$, the system $\mathcal{Q_B}$ becomes $\varrho^{\langle \epsilon|_A}_{B}(Q) = |\chi_{\epsilon}\rangle \langle \chi_{\epsilon}|_B$ with probability $|\alpha|^2[\text{cos}^2(\frac{\Delta\phi\tau}{2\hbar})+\text{sin}^2(\frac{\Delta\phi\tau}{2\hbar})|A^{\epsilon}_w|^2]$.

One can see that there is a weak value, which is $A^{\epsilon}_w=\frac{\langle \epsilon|Z|+\rangle}{\langle \epsilon|+\rangle}=\frac{\epsilon+\sqrt{1-\epsilon^2}}{\epsilon-\sqrt{1-\epsilon^2}}$. If the coupling strength of  interaction of gravity is $0$, that is $\Delta\phi=0$. In this case, the final state of system $\mathcal{Q_B}$ is $|+\rangle$ , which do not carry any weak-value information.  That is there is no entanglement between these two quantum system.  On the contrary, when the coupling strength is not 0 but small, that is, there is a gravitationally induced phase, the final state of system $\mathcal{Q_B}$ becomes  $|\chi_{\epsilon}\rangle \approx  \frac{1}{[1+|\frac{\Delta\phi}{2\hbar}A^{\epsilon}_w|^2]^{\frac{1}{2}}}(|+\rangle+i \frac{\Delta\phi}{2\hbar}A^{\epsilon}_w |-\rangle)$.  One can see that the quantum state of $\mathcal{Q}_B$ is depend on the weak value $A^{\epsilon}_w$. 
Suppose that we measure an observable $\hat{\Pi_2}=|\chi_{\epsilon}\rangle \langle \chi_{\epsilon}|$ on the system $\mathcal{Q}_B$. The observable of the displacement of $\Pi_2$ is the expectation value of the final state minus the expectation value of the initial state ($|+\rangle$) of $\mathcal{Q}_B$, and we get
\begin{equation}
	\begin{split}
		\langle \Delta \hat{\Pi_2}\rangle_Q&=\text{Tr}(\Pi_2 \varrho^{\langle \epsilon|_A}_B)- \text{Tr}(\Pi_2|+\rangle \langle +|)\\
		&= V^Q_{\Pi_2}(\varrho^{\langle \epsilon|_A}_B)-\frac{1}{[1+|\frac{\Delta\phi}{2\hbar}A^{\epsilon}_w|^2]}\\
		&	=1-\frac{1}{[1+|\frac{\Delta\phi}{2\hbar}A^{\epsilon}_w|^2]}.
	\end{split}
\end{equation}

One can see that if  $\frac{\Delta\phi}{2\hbar}A^{\epsilon}_w=1$, $\langle \Delta \hat{\Pi_2}\rangle_Q$=$\frac{1}{2}$. Even though the heralded probability $2|\alpha|^2=1-2\epsilon\sqrt{1-\epsilon^2}$ is small, after many runs of experiment, we can still observe a clear shift of the quantum state, which is a signal of entanglement between two system introduced by gravity (average shift will be $0$ without entanglement generation).  

For example, in the Ref \cite{VV}, the shift of quantity about entanglement is $p_1=sin^2(\frac{\Delta\phi}{2\hbar})\approx \frac{\Delta \phi^2}{4\hbar^2}$ in the case of weak coupling, where $p_1$ is the the probabilities  for the
mass to emerge on path 1 (R).  For showing the clear enhancement of our scheme, we give a simple example. 
If we set $A_w^{\epsilon}=10^4$ and $\frac{\Delta \phi^2}{4\hbar^2}=10^{-4}$, we have $	\langle \Delta \hat{\Pi_2}\rangle_Q=\frac{1}{2}$, while $p_1\approx 10^{-4}$. That is we can enhance the sensitivity and resolution for detect the quantum gravity $0.5\times10^4$ times by using weak-value amplification scheme. However, in above example, the steered probability becomes $p=2\times10^{-8}$ in weak-value scheme. 
Fortunately,  for existing quantum technologies, the frequency of experiments can reach MHz and beyond (suppose the time of gravitational interaction within microseconds). That is we may have $10^6$ runs in one second .  The total run of experiments is about $p\times10^6\times3600\times24=864$ each day. This is enough for us to achieve an accurate experimental estimation. Given a resolution of measurement, we achieved $X=10^4$ saving for coupling strength of gravity. In other words, we can reduce the mass of two systems by 10 times, and shorten the coupling time by 100 times. This is a very experiment-friendly scheme, which increases the feasibility of testing gravitationally induced entanglement by using existing technology.

\section*{Appendix C: Observing quantum gravity using   Limited resolution of measurement device}

In a von Neumann–type measurement, the pointer is shifted proportional to the eigenvalues of the measured observable 
\begin{equation}
	|\psi\rangle \otimes |\phi(q)\rangle \rightarrow \sum_a \langle a|\psi\rangle \cdot |a\rangle \otimes |\phi (q-g_0a)\rangle,
\end{equation}
where $\Psi$ and $\phi(q)$ are the initial states of system and probe, respectively,  the index $a$ refers to the eigenbasis of the observable, $q$ is the position of the probe, and $g0$ is a coupling constant. The outcome of the measurement is then provided by reading the position of the probe. 

In a ideal projective measurement the probe's initial state is narrower than the distance between the eigenvalues, i.e., $\langle \phi(q-a)|\phi(q-a')\rangle =\delta_{aa'}$, hence, reading the probe's
position provides full information of the measured physical quantity and collapses the system into the corresponding eigenstate of the observable.
However, it has been show that the ideal projective measurements can not be implemented in experiments since they need infinite resource costs \cite{projective}.
Therefore, the resolution of measurement devices are always limited, that is $|\langle \phi(q-a)|\phi(q-a')\rangle|^2 =\gamma\ne0$.
This is the noise come from the measurement process.

In the previous section, we consider two quantum system in initial state $|+\rangle|+\rangle$ interact each other by gravity, and one of them poselected to a almost completely orthogonal state $|\epsilon\rangle=\epsilon|0\rangle-\sqrt{1-\epsilon^2}|1\rangle$, where $\epsilon$ is close to $\frac{1}{\sqrt{2}}$. This postseleted operation is exactly limited by the resolution of measurement $\gamma$. So the minimal overlap  $ \langle \epsilon|+\rangle=\sqrt{\gamma}$, which determines the upper limit of the weak value, $Max(A^{\epsilon}_w)\approx\frac{\sqrt{2}}{\sqrt{\gamma}}$.

Now let consider the case that the square of effective coupling strength $\frac{\Delta \phi^2}{4\hbar^2}= \gamma$.
That is one can not measure the entanglement using traditional entanglement witness methods  \cite{Bose,VV}, since the signal of gravitationally induced entanglement is covered by the noise of measurement device.  In this case, weak-value based scheme is still work. One can obtain obvious signal of gravitationally induced entanglement $\langle \Delta \hat{\Pi_2}\rangle_Q=\frac{2}{3}\gg \gamma$.

\section*{Appendix D: Observing gravitationally induced entanglement with decoherence}
In fact, the decoherence is exist in experiment. The longer time in single run, the more decoherence.  Besides, the evolution of desired initial quantum state is not ideal because the system is inevitably coupled with the environment. Let us consider an environment-induced decoherence model (other decoherence models are out of our analysis) for the system $\mathcal{Q_A}$ and $\mathcal{Q_B}$. 
Without loss generality, we consider  the action of environment as a partially depolarizing channel, which is given as 
\begin{equation}
	N_E(\varrho)=(1-q)\varrho+q\frac{I}{d},
\end{equation}
where $q$ corresponds to the degree of a system that has been decohered.  
Therefore, the final state before measurement has a mathematical form
\begin{equation}
	\varrho_{AB}= (1-q)|\Psi\rangle \langle \Psi|+q\frac{ I_A\otimes I_B}{4},
\end{equation}
where $|\Psi\rangle=\alpha |\epsilon\rangle_A \otimes |\tilde{\chi}_{\epsilon}\rangle_B + \beta |\epsilon^{\perp}\rangle_A \otimes |\tilde{\chi}_{\epsilon^{\perp}}\rangle_B $. Therefore, we can get four steered states 
\begin{equation}\label{eq4Noise}
	\begin{split}
		\tilde{\varrho}^{\langle 0|_A}_{B}(Q) &=\frac{(1-q)}{2}  |\phi_+\rangle \langle \phi_+|_B+\frac{q}{2}I_B,\\
		\tilde{\varrho}^{\langle 1|_A}_{B}(Q)  &=\frac{(1-q)}{2} |\phi_{-}\rangle\langle \phi_{-}|_B+\frac{q}{2}I_B,\\
		\tilde{\varrho}^{\langle \epsilon|_A}_{B}(Q)  &=(1-q)|\alpha|^2 \text{Tr}(|\tilde{\chi}_{\epsilon}\rangle \langle \tilde{\chi}_{\epsilon}|) |\chi_{\epsilon}\rangle \langle \chi_{\epsilon}|_B+\frac{q}{2}I_B,\\
		\tilde{\varrho}^{\langle \epsilon^{\perp}|_A}_{B}(Q)  &=(1-q)|\beta|^2 \text{Tr}(|\tilde{\chi}_{\epsilon^{\perp}}\rangle\langle \tilde{\chi}_{\epsilon^{\perp}}|) |\chi_{\epsilon^{\perp}}\rangle\langle \chi_{\epsilon^{\perp}}|_B+\frac{q}{2}I_B.\\
	\end{split}
\end{equation}
Further, we can project quantum states of $\mathcal{Q_B}$ to $|\phi_+\rangle, |\phi_{-}\rangle, |\chi_{\epsilon}\rangle, |\chi_{\epsilon^{\perp}}\rangle$  with probabilities
\begin{equation}\label{probb}
	\begin{split}
		&\{p^{(0,\phi_+)}, p^{(1,\phi_-)},p^{(\epsilon,\chi_{\epsilon})}, p^{(\epsilon^{\perp},\chi_{\epsilon^{\perp}})}\}  \\
		& = \{\frac{1}{2}-\frac{q}{4},\frac{1}{2}-\frac{q}{4},(1-q)|\alpha|^2 \text{Tr}(|\tilde{\chi}_{\epsilon}\rangle \langle \tilde{\chi}_{\epsilon}|)+\frac{q}{4},(1-q)|\beta|^2 \text{Tr}(|\tilde{\chi}_{\epsilon^{\perp}}\rangle\langle \tilde{\chi}_{\epsilon^{\perp}}|) +\frac{q}{4}\},
	\end{split}
\end{equation}

Further, the visibility of measurement for quantum mediator are 
\begin{equation} \label{vis}
	\begin{split}
		V^{noise}_{\Pi_0} (\varrho^{\langle 0|_A}_{B})  &=   \frac{1}{1+q}   ;          \\
		V^{noise}_{\Pi_1}  (\varrho^{\langle 1|_A}_{B})   &= \frac{1}{1+q}       ;              \\
		V^{noise}_{\Pi_2}  (\varrho^{\langle \epsilon|_A}_{B}) &=     1-\frac{q}{2\text{Tr}(\tilde{\varrho}^{\langle \epsilon|_A}_{B})}   ;        \\
		V^{noise}_{\Pi_3} (\varrho^{\langle \epsilon^{\perp}|_A}_{B}) &=1-\frac{q}{2\text{Tr}(\tilde{\varrho}^{\langle, \epsilon^{\perp}|_A}_{B})}    
	\end{split}
\end{equation}
where $i=0,1,2,3$. Similarly, we can construct a separable state \eqref{C} model for comparison. If $q$ is small, i.e.  so that $V^{noise}_{\Pi_2}  (Q) >V^{noise}_{\Pi_2}  (C) $, we can still obtain the amplified  signal of entanglement. Therefore, our weak entanglement criterion is applicable to some kinds of  mixed states.

\section*{Appendix E :  atomic interferometers with a harmonic oscillator}

Here we analyze a possible experimental proposal based on atomic interferometers with a harmonic oscillator \cite{PRX}. We analyze how to apply our weak entanglement criterion to this scenario. Consider a harmonic oscillator $\mathcal{Q_A}$ (a mechanical resonator) coupled to a two-state system $\mathcal{Q_B}$ (an atom trapped in a double-well potential).  Since $\mathcal{Q_B}$  is a qubit state, one can set the position operator of the atom $\mathcal{Q_B}$ to the Pauli matrix $\sigma_z$ with the eigenstates $|L\rangle$ and $|R\rangle$, which represent the location of atom occupying, respectively. The gravitationally driven Hamiltonian of these two system is given as ($\hbar=1$) \cite{PRX}
\begin{equation}
	H=w a^{\dagger}a+ g(a+a^{\dagger})\sigma_z,
\end{equation}
where $w$, $a^{\dagger}$ and $a$ denote the frequency, creation and annihilation operators of the harmonic oscillator, respectively. The coupling coefficient $g$ correspond to the gravitational interaction between atom and oscillator, satisfying $g\ll w$. 
Up to a global phase, the time-evolution operator can be rewritten as 
\begin{equation}
	U(t)=D^{\dagger }(\sigma_z \lambda) e^{-iwa^{\dagger}at}D(\sigma_z \lambda), 
\end{equation}
where $D(\zeta)\equiv \text{exp}\{\zeta a^{\dagger} -\zeta^{*} a \}$ is the usual displacement operator and $\lambda =\frac{g}{w}$.
Consider the oscillator is initialized in its ground state $|0\rangle_A$ and the atom is in the superposition of $|L\rangle_B$ and $|R\rangle_B$. After the time evolution, the composite quantum state becomes
$	|\Psi (t) \rangle = U(t) |0\rangle_A \otimes \frac{1}{\sqrt{2}}(|L\rangle_B +|R\rangle_B)
=\frac{1}{\sqrt{2}}  ( |\eta \rangle_A \otimes |L\rangle_B  +  |-\eta \rangle_A \otimes |R\rangle_B ),$
where the evolved states of the oscillator are coherent states $|\pm \eta \rangle_A=D(\pm\lambda(e^{-iwt}-1))$.
If we implement the Hadamard gate to  the two-level system $\mathcal{Q_B}$, we have 
\begin{equation} \label{cat}
	|\Psi (t) \rangle =\frac{1}{2c_{+}}  |\text{cat}_{\eta_+}\rangle_A \otimes |L\rangle_B +\frac{1}{2c_{-}} |\text{cat}_{\eta_-}\rangle_A \otimes |R\rangle_B,
\end{equation}
where $ |\text{cat}_{\eta_+}\rangle_A=c_{+}(|\eta \rangle_A+|-\eta \rangle_A)$ and  $|\text{cat}_{\eta_-}\rangle_A=c_{-}(|\eta \rangle_A-|-\eta \rangle_A)$  are the Schr$\ddot{\text{o}}$dinger's cat state with  $c_{+}=\frac{1}{\sqrt{2(1+e^{-2|\eta|^2})}}$ and $c_{-}=\frac{1}{\sqrt{2(1-e^{-2|\eta|^2})}}$.  Since $ \langle  \text{cat}_{\eta_-}|\text{cat}_{\eta_+}\rangle=0$, one can address Eqn. \eqref{cat}  as two-qubit entangled state.  However, the entanglement $|\Psi (t) \rangle $ is very weak ($\frac{1}{2c_{-}}$ is small) due to the fact that $|\eta|$ is very small (determine by gravitational interaction).
Similar to Eqn. \eqref{gravitystate}, one may choose a suitable basis for $\mathcal{Q_A}$ ($\mathcal{Q_B}$) to a new form of $|\Psi (t) \rangle$, which has the amplified signal.  For example, let's expand $\mathcal{Q_A}$ to the basis $\{| v\rangle =\text{sin}(\theta)|\text{cat}_{\eta_+}\rangle +\text{cos}(\theta)|\text{cat}_{\eta_-}\rangle, |v^{\perp}\rangle =\text{cos}(\theta)|\text{cat}_{\eta_+}\rangle - \text{sin}(\theta)|\text{cat}_{\eta_+}\rangle\}$ with $\theta \ll 1$, and one has  
\begin{equation} \label{cat2}
	\begin{split}
		|\Psi (t) \rangle = |v\rangle_A \otimes (\frac{\langle v | \text{cat}_{\eta_+}\rangle}{2c_{+}}|L\rangle_B+\frac{\langle v | \text{cat}_{\eta_-}\rangle}{2c_{-}}|R\rangle_B)  \\
		+ |v^{\perp}\rangle_A \otimes (\frac{\langle v^{\perp} | \text{cat}_{\eta_+}\rangle}{2c_{+}}|L\rangle_B+\frac{\langle v^{\perp} | \text{cat}_{\eta_-}\rangle}{2c_{-}}|R\rangle_B).
	\end{split}
\end{equation}
One can see that compared to the components of  $|L\rangle_B$, the component of $|R\rangle_B$ is enlarged when $\mathcal{Q_A}$ is projected to $|v\rangle_A$. If $\theta $ is small enough, one may has $\frac{\langle v | L\rangle}{2c_{+}}\sim \frac{\langle v | R\rangle}{2c_{-}}$ such that the component of $|R\rangle_B$ in the steered quantum state of $\mathcal{Q_B}$  is boosted. According to weak entanglement criteria we proposed, this basis is the most significant ingredient to amplify the entangled signal.  Similarly, another measurement basis needs to be selected, which may be $\{| +\rangle =\frac{1}{\sqrt{2}}|\text{cat}_{\eta_+}\rangle +\frac{1}{\sqrt{2}}|\text{cat}_{\eta_-}\rangle, |-\rangle =\frac{1}{\sqrt{2}}|\text{cat}_{\eta_+}\rangle -\frac{1}{\sqrt{2}}|\text{cat}_{\eta_-}\rangle\}$. A random selection of these two measurement bases yields the probability distribution and measurement visibility of system $\mathcal{Q_B}$, which allows us to witness the gravitationally induced entanglement. In real scenarios, the resonator $\mathcal{Q_B}$  may be a thermal state close to the ground state.  We analyze this case  in Appendix E.

In above analysis, we assume that the oscillator is initialized in its ground state $|0\rangle$. In a realistic implementation,  due to the finite temperature (may be nK), the oscillator instead starts in a mixed state, such as a thermal state, donated as $\varrho_{th}=\int d^2 \zeta  \frac{1}{\pi\bar{n}}e^{-|\zeta|^2/\bar{n}} |\zeta\rangle \langle \zeta |$. In this case, the evolving state of two system becomes                                
\begin{equation} 
	\varrho_{AB}=\int d^2 \zeta  \frac{1}{\pi\bar{n}}e^{-|\zeta|^2/\bar{n}} |\Psi_\zeta\rangle \langle \Psi_\zeta|,
\end{equation}
where $|\Psi_\zeta\rangle=\frac{1}{2}  ( |\zeta+\eta \rangle_A + |\zeta-\eta \rangle_A ) \otimes  |L\rangle_B + \frac{1}{2}  (  |\zeta+\eta \rangle_A-|\zeta-\eta \rangle_A  )\otimes |R\rangle_B$. Obviously, if $\zeta=0$, $|\Psi_\zeta\rangle$ reduce to Eqn. \eqref{cat}.
Here we will be concerned only with the projective measurements $\langle v|_A$ (corresponds to amplified entanglement), since the measurements of the other bases are trivial. The conditional state of the atom becomes 
\begin{equation} 
	\varrho^{\langle v|_A}_B= \frac{\text{Tr}_A( |v\rangle \langle v|_A  \otimes I_B \varrho_{AB})}{\text{Tr}( |v\rangle \langle v|_A \otimes I_B \varrho_{AB})},
\end{equation}
leading the measurement visibility 
\begin{equation} 
	V_{\Pi_{\mu}}(\varrho^{\langle v|_A}_B)=\text{Tr} (\Pi_{\mu}\varrho^{\langle v|_A}_B)
\end{equation}
where $\Pi_{\mu}=|\mu \rangle \langle \mu|$ and $|\mu\rangle =(\frac{\langle v | \text{cat}_{\eta_+}\rangle}{2c_{+}}|L\rangle_B+\frac{\langle v | \text{cat}_{\eta_-}\rangle}{2c_{-}}|R\rangle_B)/\sqrt{|\frac{\langle v | \text{cat}_{\eta_+}\rangle}{2c_{+}}|^2+|\frac{\langle v | \text{cat}_{\eta_-}\rangle}{2c_{-}}|^2}$.  
When the thermal state is very close to the ground state, 
one may has $\frac{V_{\Pi_{\mu}}(\varrho^{\langle v|_A}_B)- V^C_{\Pi_{\mu}}(\varrho^{\langle v|_A}_B)}{\gamma}=k>1$, where $\gamma$ is measurement sensitivity.  Therefore we may still achieve $k$-fold magnification compare to the usual ones.


\begin{thebibliography}{99}
\bibitem{note1}
	It should be noted that, traditionally, the general weak value derivation is based on the Taylor approximation under weak coupling, because not all the evolution of the Hamiltonian can be accurately expanded.
\bibitem{note}
	Note that if the logical $|0\rangle$  and $|1\rangle $ are redefined, any two-qubit pure state has a mathematical form such as $|\Psi\rangle$, so our analysis later will not lose the generality.
\end{thebibliography}
\end{document}